\documentstyle{article}

\newcommand{\mathbf}{\bf}

\begin{document}

\begin{center}
{\huge\bf The Canonical Flux Quantization and IQHE}
\end{center}

\vspace{1cm}
\begin{center}
{\large\bf
F.GHABOUSSI}\\
\end{center}

\begin{center}
\begin{minipage}{8cm}
Department of Physics, University of Konstanz\\
P.O. Box 5560, D 78434 Konstanz, Germany\\
E-mail: ghabousi@kaluza.physik.uni-konstanz.de
\end{minipage}
\end{center}

\vspace{1cm}

\begin{center}
{\large{\bf Abstract}}
\end{center}
It is shown that the canonical flux quantization, which is  
described by the uncertainty relation on the phase space of the flux  
system, can result in the quantization of Hall-measures. Further it  
is shown that the polarization of this phase space, which is  
necessary for its quantization, results in the vanishing of  
longitudinal resistivity and conductivity. The equivalence between  
this approach and the topological approach to QHE is also discussed.
\begin{center}
\begin{minipage}{12cm}

\end{minipage}
\end{center}

\newpage
From the preference of edge currents in IQHE and the restriction of  
electronic currents in the  theory of flux quantization to the  
boundary regions of superconducting ring, one deduces that there  
must be a connection between the IQHE and Flux quantization or  
superconductivity. Hence there is also hints about such a connection  
in case of FQHE \cite{Laughlin}. We discussed such a connection  
already in a previous paper \cite{sucq}. Here we will show that one  
can derive a possible background for the IQHE from the flux  
quantization.

It is well known that the flux quantization $\oint e A_m dx^m =  
\int \int e F_{mn} dx^m \wedge dx^n = \Phi = {\mathbf Z} h ; m, n =  
1, 2$ is an experimentally varified fact, however the conventional  
theory of flux quantization does not match the canonical  
quantization. Nevertheless in principle not only the flux  
quantization but also the QHE ,as quantum theories, should be  
described by a canonically quantized theory.

We showed recently that the flux quantization can be described as  
the canonical quantization of $\oint A_m dx^m$ on the ${\{ A_m ,  
x^m}\}$-phase space by the commutator postulate $e [\hat{A}_m ,  
\hat{x}_m] = -i \hbar$. Thus, there must be an equivalent  
uncertainty relation that can appear in two dependent forms:

\begin{eqnarray}
&&e \Delta A_m \cdot \Delta x_m \geq \hbar \nonumber\\
&&e B \cdot |\Delta x_m |\cdot |\Delta x_n \epsilon_{mn}| \geq \hbar \ ,
\end{eqnarray}

which are related with each other by the Landau gauge $A_m := B x_n  
\epsilon_{mn}$ \cite{mein}.

Recall that the canonical flux quantization with respect to the  
$\int \int F_{mn} dx^m \wedge dx^n = e B \int \int dx^m \wedge dx^n  
= {\mathbf Z} h$ action can also be described by $e B [\hat{x}_m ,  
\hat{x}_n ] = -i \hbar \epsilon_{mn}$ commutator which is related  
with the second form of the uncertainty relation (1) and which is  
known as the non-commutativity of the relative coordinates of the  
cyclotron motion of electron \cite{landau}. This commutator is  
related with the above one by the Landau gauge.

If we consider the minimum value of position uncertainty $(\Delta  
x_m)_{(min)} = l_B$, then we obtain from the second uncertainty  
relation the defining relation of the magnetic length: $l_B ^2 :=  
\displaystyle{\frac{\hbar}{eB}}$. Thus, one has also the uncertainty  
equations:

\begin{equation}
e \Delta A_m \cdot l_B = e B \cdot l_B ^2 =\hbar
\end{equation}

Here we will show that the uncertainty relations (1) or (2) can  
also define the quantization of the Hall- conductivity or  
resistivity.
Recall also that the mentioned commutator $[\hat{x}_m , \hat{x}_n ]  
= -i l_B ^2 \epsilon_{mn}$ is an example of the non-commutative  
geometry of configuration space of cyclotron motion, hence there is  
also a quantization of the Hall-measures according to the  
non-commutative geometry \cite{bellisard}.

We will show also that, if the flux $\Phi := \oint e A_m dx^m =  
\int \int F_{mn} dx^m \wedge dx^n$ is considered as a canonical  
action functional on the phase space of the related system, then its  
variation results in the correct equations of motion of photons.  
Furthermore, it will be shown that according to the "geometric"  
quantization of flux, the necessary polarization of the phase space  
of flux results in the vanishing of longitudinal resistivity and  
conductivity in QHE \cite{geoq}. In other words, the canonical  
quantization of flux should explain both: The quantization of Hall-  
measure and the vanishing of longitudinal components, in a  
consistent manner.

\bigskip
To discuss the quantization let us first consider that in the Ohm's  
equation the Hall resistivity is given as the proportionality  
factor $\rho_H = \displaystyle{\frac{\partial_t A_m}{ne \partial_t  
x_n}} \epsilon_{mn}$ or $\rho_H = \displaystyle{\frac{\partial  
A_m}{ne \partial x_n}} \epsilon_{mn}$ which is equivalent to its  
semi-classical
definition $\rho_H := \displaystyle{\frac{B}{ne}}$, where $n$ is  
the global density of electrons on the QHE sample. These both  
definitions are however semi-classical definitions as one expects  
from the usual (semi-classical) Hall-effect. Whereas in the  
quantized case, i. e. in QHE, the relation between $A_m$- and $x_m$  
expectation values should be governed either by the expectation  
values of the above discussed commutators or by the equivalent  
uncertainty relation (1). In other words, in such a quantum case,  
one has to do always with dependent variations of $A_m$ and $x^m$ so  
that: $\partial A_m \geq \Delta A_m$ and $\partial x_m \geq \Delta  
x_m$ which are related by (1) or (2).

Therefore, in the QHE case the quantized Hall resistivity should be  
given instead of $\rho_H = \displaystyle{\frac{\partial A_m}{ne  
\partial x_n}} \epsilon_{mn}$, consistently, by its quantized  
version where the differentials $\partial A_m$ and $\partial x_m$  
are replaced by the related finite values of  $\Delta A_m$ and  
$\Delta x_m$ uncertainities:

\begin{eqnarray}
&&\rho_H ^Q = \displaystyle{\frac{\Delta A_m}{ne \Delta x_n}}  
\epsilon_{mn}\ ,\nonumber\\
&&e \Delta A_m \cdot \Delta x_m = e B \Delta x_m \Delta x_n \geq  
\hbar \ ,
\end{eqnarray}

which means that the uncertainty relation should be fulfilled by  
the $\rho_H ^Q$ relation.

If we replace $\Delta A_m$ in the $\rho_H ^Q$ relation according to  
the first uncertainty relation (1), then we arrive at:

\begin{equation}
\rho_H ^Q = \epsilon_{mn} \displaystyle{\frac{\hbar}{ne^2 \Delta  
x_m \cdot \Delta x_n}}
\end{equation}

Recall, that the classical and semi-classical relations are given  
in terms of infinitesimals, e. g. $\partial x^m$ which should tend  
to zero. Whereas, the quantum relations are in terms of finite  
quantities, e. g. $\Delta A_m$ and $\Delta x^m$ which are prevented  
to become zero {\it in the quantum cases} according to the  
uncertainty relations.

Now the most closest form of the quantum relation (4) to its  
differential or local definition: $\rho_H =  
\displaystyle{\frac{\partial A_m}{ne \partial x_n}} \epsilon_{mn}$  
is achievable for the most minimal possible values of $\Delta A_m$  
and $\Delta x_n$. Nevertheless, this can be achieved, in view of the  
dependency $A_m = \epsilon_{mn} B \cdot x_n$, also if only $\Delta  
x_n = \Delta x_m$ is replaced by its most minimal value which is  
equal to the magnetic length $l_B$. Because, for a given value of  
$B$, the $(\Delta x_n)_{(min)} = l_B$ is the smallest possible  
length which is available quantum theoretically for the cases under  
consideration and so it is the smallest variation for $x_m$ under  
the given quantum conditions. Consistently, also for the ideal case  
of QHE one has the edge currents which flow within a distance from  
the edge of sample which is equal to the magnetic length, i. e.  
$(\Delta x_m)_{(min)} = l_B$ \cite{KL}. Moreover, one has for this  
same ideal case under quantum conditions (QHE) the uncertainty  
equation (2) where $|\Delta x_m|_{(min)} = |\epsilon_{mn} \Delta  
x_n|_{(min)} = l_B$. Using this in the relation (4) one obtains the  
quantized value of the Hall-resistivity:

\begin{equation}
\rho_H ^Q = \displaystyle{\frac{h}{\nu e^2}} \qquad ,
\end{equation}

where $\nu := 2 \pi n l_B ^2$ is as usual the filling factor.

Thus the most closest form of the quantum theoretically allowed  
definition of $\rho_H ^Q$ to its infinitesimal definition $\rho_H =  
\displaystyle{\frac{\partial A_m}{ne \partial x_n}} \epsilon_{mn}$,  
which is also appropriate for the QHE case, is given by the relation  
(5).

Therefore, the flux quantization on the QHE sample which is  
described by the uncertainty equation (2) can describe also the  
quantization of the Hall-resistivity, i. e.: $\rho_H =  
\displaystyle{\frac{\partial A_m}{ne \partial x_n}} \epsilon_{mn}  
\rightarrow \rho_H ^Q = \displaystyle{\frac{h}{\nu e^2}}$ in the  
above discussed manner \cite{alternat}.

\bigskip
This approach is, despite of its finite character according to  
$\Delta x^m$ and $\Delta A_m$, also a topological approach, because  
on the one hand the area of sample under QHE-conditions is a  
multiple of the minimal quantum cell area: $ 2 \pi l_B ^2$. On the  
other hand, in the topological approach to QHE \cite{thouless} the  
quantized Hall-measures for the sample are given as the surface  
integral of the applied magnetic field, which is as the first Chern  
number on the sample manifold a topological invariant. Now for the  
applied  constant magnetic field $B$ it is obvious that this  
invariant is given by quantized flux $e B \cdot S = \tilde {N} h$  
where $S$ is the total area of the sample and equal to a multiple of  
the above mentioned quantum cell area. Thus, it is this  
multiplicity of the total area of sample under quantum conditions,  
i. e. $S = \tilde{N} \cdot 2\pi l_B ^2$ which results in the  
quantization of the Hall-measures.

Recall that the same area is also involved in the $N = n \cdot S$
\cite{alternat}.

To see the equivalence with the topological approach in a direct  
way, let us perform the surface integral of the above definition $  
ne \rho_H = \displaystyle{\frac{\partial A_m}{\partial x_n}}  
\epsilon_{mn}:= B$ over the area of sample: $\int \int ne \rho_H =  
\int \int B$. It results in view of the local constancy of $\rho_H$  
and for a constant $B$ in the quantized Hall resistivity: $ Ne^2  
\rho_H ^Q = \tilde{N} h$ or $\rho_H = \displaystyle{\frac{h}{\nu  
e^2}}$ \cite{alternat}.

In summary: the flux quantization defines according to the relation  
(2) for a given $B$ a quantum measure for the $2-D$ area which is  
given by $ 2\pi l_B ^2$, such that the relevant "quantum"  
topological invariants , e. g. the area, are multiples of it. This  
is in accordance with the usual quantum cell decomposition of a  
quantized phase space, which is represented by the uncertainty  
relation.

\bigskip
Now we show that the flux action functional $S = \oint e A_m dx^m =  
\int \int F_{mn} dx^m \wedge dx^n$  results in the true equation of  
motion for the electromagnetic potential, i. e. for the photon  
field with its true two degrees of freedom.

Herefore, recall first that the mentioned action functional has to  
be considered on the phase space of the system where its canonical  
conjugate variables are given by the set ${\{A_m, x^m}\}$. However,  
in view of the fact that $A_m$ depends on $x^n$ by $A_m = B x^n  
\epsilon_{mn}$, the variation of action $\delta S$ needs to be  
considered only with respect to the variation of $\delta x^n$,  
because the variation $\delta A_m$ is proportional to $\delta x_n$.  
Nevertheless in view of the action preserving canonical  
transformations on the phase space, $x^m$ as variables on the above  
phase space are in general functions of $x^n$, therefore one has to  
consider $dx^l = \displaystyle{\frac{\partial x^l}{\partial x^m}}  
dx^m$. To be precize, let us mention that the situation is the  
following: In the above phase space the varibles ${\{A_m, x^m}\}$  
has to be considered as functions of $x^l$, whereby the system is  
constrained by $A_m = B x^n \epsilon_{mn}$.

Let us consider now the $S = \int \int F_{mn} dx^m \wedge dx^n$ and  
look for the Euler-Lagrange equations which results from a  
variation of this action with respect to the variation $\delta x^n$.  
The mentioned Euler-Lagrange equations  
$\displaystyle{\frac{\partial L}{\partial x^m}} = \partial _n  
{\frac{\partial L}{\partial \partial_n x^m}}$ results in

\begin{equation}
\partial^n \partial_n A_m - \partial_m \partial^n A_n = 0 \ ,
\end{equation}

which is the usual vacuum equations of motion for $A_m$ potential,  
here in two dimensions. If one introduces the Lorentz gauge  
$\partial^n A_n = 0$, then one obtains the Laplace equation  
$\partial_n \partial^n A_m = 0$ as the equations of motion of  
electrodynamics. Recall that the other part of equations of motion  
in electrodynamics $d F = d^2 A = 0$ are identities.

The same result is also achievable, if one uses the $S = \oint A_m  
dx^m$ form of the same action functional. Here, one should consider  
fairly $S = \displaystyle{\frac{1}{2}} \oint (A_m dx^m - x^m dA_m)$  
\cite{action}. Thereafter, one obtains from the above Euler-Lagrange  
equations, the equations
$d A_m = \partial_m A_m$ which is fulfilled only if both sides are  
equal to zero, whereby $d A_m := \partial_n A_m dx^n \epsilon_{mn}$.  
Now these together, i. e. $ d^{\dagger} A = \partial_m A_m = 0$ and  
$d A = dA_m dx^m = \partial_n A_m dx^n \wedge dx^m = 0$, give the  
same Laplace equation $ (d^{\dagger} d + d d^{\dagger}) A =  
\partial_n \partial^n A_m = 0$ as above.

We show further that the necessary polarization in the phase space  
space of flux quatization, if it is considered for the QHE case,  
results in the vanishing of the longitudinal resistivity or  
conductivity \cite{pure}.

\bigskip
The concept of polarization is an essential tool of the  
quantization theory \cite{geoq}. It describes the well known  
circumstance that the state function, which fulfils the quantum  
equation of motion, is beyond its time dependency only a function of  
the half of the phase space variables. This is known usually as the  
representation of the wave function, so that it should be either in  
the position or in the momentum representation. Thus, the  
polarization in the phase space of the canonical system ${\{ p_m ,  
q^m}\}$ is described mathematically by the condition that the wave  
function $\Psi$ should fulfil either
$\displaystyle{\frac{\partial \Psi}{\partial p_m}} = 0$ or  
$\displaystyle{\frac{\partial \Psi}{\partial q^m}} = 0$ \cite  
{geoq}, \cite{gemisch}.

As already discussed the phase space variables of the flux system  
are ${\{ A_m , x^m}\} \cite{pure}$. Thus the wave function of this  
system $\psi$ should be, beyond its time dependency, only a function  
of $A_m$ or of $x^m$, i. e. either $\psi (A_m , t)$ or $\psi(x^m ,  
t)$. In other words, in the first case it should fulfil the  
polarization equation:

\begin{equation}
\frac{\partial \psi (A_m , t)}{\partial x^m} = 0
\end{equation}

However, in view of the fact that $\psi (A_m (x^n) , t)$ does not  
depend explicitely on $x^m$, but its $A_m$-variables can be in  
general functions of $x^m$, the following polarization equations  
should be solved on the integration path, where the action  
functional is defined:

\begin{equation}
{\frac{\partial \psi}{\partial A_m}}{\frac{\partial A_m}{\partial  
x^m}} + {\frac{\partial \psi}{\partial A_n}}{\frac{\partial  
A_n}{\partial x^m}} \epsilon_{mn} = 0
\end{equation}

It is known that in the flux quantization the path of integration  
in the action functional $\oint A_m dx^m$ must be chosen in a region  
where the electronic current is absent. This means that the  
electromagnetic potential {\it on this path} is a pure gauge  
potential $\breve{A}_m = \partial_m \Phi$ \cite{mein}. Therefore,  
{\it on} such an integration path the magnetic field strength should  
vanish $(F_{mn})_{(on)} = F_{mn} (\breve{A}_m) =  
\displaystyle{\frac{\partial \breve{A}_n}{\partial  
x^m}}\epsilon_{mn} = 0$. This can be understood also in the  
following way: That because the velocity operators $\hat{V}_m =  
(\hat{P}_m - e\hat{A}_m)$ which define the current must vanish on  
this path, hence the field strength defined by the commutator of  
current operators $ (F_{mn})_{(on)} = [ \hat{V}_m , \hat{V}_n ]$   
also should vanish.

Taking this into account the polarization equation (8) reduces to;

\begin{equation}
\displaystyle{{\frac{\partial \psi}{\partial  
\breve{A}_m}}{\frac{\partial \breve{A}_m}{\partial x^m}}} = 0\
\end{equation}

From the semi-classical Ohm's equations $\partial_t \breve{A}_m =
\rho_L \cdot j_m + \rho_H \cdot j_n \epsilon_{mn}$ in the boundary  
region, it results that the longitudinal resistivity should be  
defined by the ratio $\rho_L := \displaystyle{\frac{\partial_t  
\breve{A}_m}{ne \partial_t x^m}}$ or by $ ne \rho_L :=  
\displaystyle{\frac{\partial \breve{A}_m}{\partial x^m}}$.  
Therefore, in view of the necessary dependency of $\psi$ on  
$\breve{A}_m$ in this polarization direction, the polarization  
equation (9) is fulfilled only if the longitudinal resistivity  
vanishes:

\begin{equation}
 ne \rho_L = \displaystyle{\frac{\partial \breve{A}_m}{\partial x^m}} = 0
\end{equation}

Recall that, if one considers the other possible polarization  
direction $ \displaystyle{\frac{\partial \psi}{\partial  
\breve{A}_m}} = 0$ in the same phase space, then one arrives in the  
vanishing of the longitudinal conductivity $\sigma_L :=  
\displaystyle{\frac{ne \partial x^m }{\partial \breve{A}_m}}$. In  
view of the necessary independency of the quantization from the  
direction of polarization in a quantized case like QHE \cite{geoq}  
\cite{gemisch}, both of mentioned longitudinal measures $\sigma_L$  
and $\rho_L$ should vanish.

The absolute vanishing of these longitudinal measures is an  
indication of an ideal QHE, hence the exact polarization describes  
an exact quantization; Whereas in real QHE samples, in view of  
impurieties, these longitudinal measures do not vanish exactly. In  
an ideal QHE case where the quantization is exact the electronic  
currents flows within the width of magnetic length on the edge of  
sample, thus the potential uncertainty in this ideal case should be  
given by $\Delta A_m = \displaystyle{\frac{\hbar}{e l_B}}$. Whereas  
in real QHE cases the quantization or polarization is not exact and  
not only the longitudinal measures do not vanish exactly but also  
the potential uncertainty should be less than that in the ideal case  
(see also \cite{mein}).

It is also intresting to mention that the vanishing of longitudinal  
measures or the polarization of the flux/QHE phase space is  
equivalent to the Lorentz/Coulomb gauge fixing in such a $2-D$ case.

Thus, the vanishing of longitudinal resistivity or conductivity in  
QHE is a result of the necessary quantization structure, i. e. of  
the polarization on the phase space of the system. In this way the  
vanishing of longitudinal measures in QHE can be understood within  
the canonical flux quantization of the QH-system.

\bigskip
Footnotes and references


\begin{thebibliography}{100}
\bibitem{action}
In view of the fact that, according to the symplectic structure of  
the canonical phase space, the symplectic $2-form$ can be given by  
$\omega := d(p_m d q^m) = - d (q^m dp_m) =  
\displaystyle{\frac{1}{2}} d ( p_m dq^m - q^m dp_m)$. Then also the  
symplectic (closed) $2-form \ F$ on the ${\{ A_m , x^m }\}$-phase  
space of flux quantization can be given by: $F_{mn} dx^m \wedge dx^n  
= d A_m \wedge dx^m = d (A_m dx^m) = - d(x^m dA_m) =  
\displaystyle{\frac{1}{2}} d (A_m dx^m - x^m dA^m)$. Thus, the  
proper form of the closed line integral action should be given by  
the

$\displaystyle{\frac{1}{2}} \oint ( A_m dx^m - x^m dA^m)$.
\bibitem{pure}
In the first part we used merely the action $\int \int F$. In the  
polarization part we use the equivalent action $\oint A$ which is  
predistinated for quantization procedure.
\bibitem{gemisch}
Of course it is possible also to use polarizations which are linear  
complex combinations of the mentioned two ones. Nevertheless, the  
quantization should be independent of the polarization and the  
polarization should be compatible with the constraint of the theory  
( see Ref. \cite{geoq}).
\end{thebibliography}
\end{document}